\documentclass[sigconf]{acmart}
\AtBeginDocument{%
  }

\setcopyright{none}
\copyrightyear{}
\acmDOI{}
\acmConference[Preprint]{}{}{December 2025}
\acmISBN{}

\usepackage{multirow}

\begin{document}

\title{CS-Guide: Leveraging LLMs and Student Reflections to Provide Frequent, Scalable Academic Monitoring Feedback to Computer Science Students}

\author{Samuel Jacob Chacko}
\authornote{Both authors contributed equally to this research.}
\affiliation{%
  \institution{Department of Computer Science, Florida State University}
  \city{Tallahassee}
  \state{Florida}
  \country{USA}
}
\email{sj21j@fsu.edu}

\author{An-I Andy Wang}
\authornotemark[1]
\affiliation{%
  \institution{Department of Computer Science, Florida State University}
  \city{Tallahassee}
  \state{Florida}
  \country{USA}
}
\email{aawang@fsu.edu}

\author{Lara Perez-Felkner}
\affiliation{%
  \institution{College of Education, Florida State University}
  \city{Tallahassee}
  \state{Florida}
  \country{USA}
}
\email{lperezfelkner@fsu.edu}

\author{Sonia Haiduc}
\affiliation{%
  \institution{Department of Computer Science, Florida State University}
  \city{Tallahassee}
  \state{Florida}
  \country{USA}
}
\email{shaiduc@fsu.edu}

\author{David Whalley}
\affiliation{%
  \institution{Department of Computer Science, Florida State University}
  \city{Tallahassee}
  \state{Florida}
  \country{USA}
}
\email{dwhalley@admin.fsu.edu}

\author{Xiuwen Liu}
\affiliation{%
  \institution{Department of Computer Science, Florida State University}
  \city{Tallahassee}
  \state{Florida}
  \country{USA}
}
\email{xliu@fsu.edu}

\renewcommand{\shortauthors}{Chacko et al.}

\begin{abstract}
Computer Science (CS) departments often serve large student populations, making timely academic monitoring and personalized feedback difficult. While the recommended counselor-to-student ratio is 250:1, it often exceeds 350:1 in practice, leading to delays in support and interventions. We present CS-Guide, which leverages Large Language Models (LLMs) to deliver scalable, frequent academic feedback. Weekly, students interact with CS-Guide through self-reported grades and reflective journal entries, from which CS-Guide extracts quantitative and qualitative features and triggers tailored interventions (e.g., academic support, health and wellness referrals). Thus, CS-Guide uniquely integrates learning analytics, LLMs, and actionable interventions using both structured and unstructured student-generated data.
  
We evaluated CS-Guide on a four-year, \textasciitilde 20K-entry longitudinal dataset, and it achieved up to a 97\% F1 score in recommending interventions for first-year students. This shows that CS-Guide can enhance advising systems with scalable, consistent, timely, and domain-specific feedback. 

\end{abstract}




\maketitle

\section{Introduction}
Computer Science (CS) majors have grown significantly in recent years, with many departments now serving over 1,000 students~\citep{zweben2023taulbee}. Our own department experienced a 33\% increase in CS majors over the past five years, reaching \textasciitilde 1,400 students. While this growth is encouraging, it has placed immense strain on the academic support infrastructure, particularly on student advising.

National guidelines recommend a counselor-to-student ratio of 250:1~\citep{goodman2018ecological}, yet many institutions regularly exceed 350:1~\citep{schoolcounselor2025}. Likewise, our department frequently operates at a ratio greater than 350:1, leading to waiting times of up to weeks for student appointments and a severely diminished capacity for timely feedback and intervention. These conditions undermine student success and contribute to counselor burnout, turnover, and instability in support services.

Our experience has demonstrated the need for scalable support mechanisms to alleviate counselors’ workloads while delivering timely, personalized guidance to students.

Recent advancements in Large Language Models (LLMs)~\citep{radford2018improving} offer a promising opportunity. LLMs have demonstrated the ability to extract structured information from unstructured text, generate human-like responses, and assist in domain-specific reasoning. In this work, we explore whether LLMs can be leveraged to support academic monitoring by providing frequent, individualized feedback to CS students.

We present CS-Guide, a framework that uses LLMs to analyze weekly student-submitted grades and reflective journals. CS-Guide combines qualitative and quantitative features to recommend tailored interventions, including reminders, study tips, academic support, and personal referrals. Beyond offering timely feedback, CS-Guide can help ensure consistency in interventions, serve as a second opinion for counselors, and accelerate the onboarding of new advising staff by capturing domain-specific knowledge.

This work explores whether LLMs can be used to build an academic monitoring feedback framework, CS-Guide, that enables scalable, timely student support. The successful deployment of CS-Guide offers several key benefits:
\begin{itemize}
    \item \textbf{Prompt, Scalable Support}: CS-Guide enables weekly feedback based on students' self-reported grades and experiences. By considering major deadlines (e.g., drop/withdrawal), major-related policies (e.g., minimum GPA), and personal circumstances (e.g., prolonged illness), CS-Guide delivers timely support before counseling appointments. Students are better informed, and counselors can work with students on the interventions that may have greater consequences (e.g., deferring graduation).
    \item \textbf{Consistency in Recommendations}: Different counselors may suggest diverging strategies for similar student situations. For example, one may advise dropping a course to preserve GPA, while another might suggest switching to a degree program with lower GPA requirements to enable graduation without course withdrawal. CS-Guide offers consistent, data-backed recommendations, giving both students and counselors a reference point for intervention planning.
    \item \textbf{Retention of Institutional Knowledge and Advising Experience}: New counselors may take time to internalize program-specific nuances, such as course sequencing, degree rules, or drop/withdrawal impacts. CS-Guide captures past advising decisions and domain heuristics, supporting onboarding counselors with better continuity during staff transitions.
\end{itemize}

\section{Research Challenges and Constraints}
We need to overcome certain barriers and constraints before testing the feasibility of applying LLMs in the domain of academic monitoring.  The first is to obtain data and domain-specific knowledge for training LLMs.  The second involves constraints due to the research environment.

\subsection{Data Collection}
In Y (year anonymized), we received a grant from the National Science Foundation to fund multi-year scholarships to freshmen cohorts to promote CS majors.  The recipients must be U.S. citizens/permanent residents with academic ability and financial need as defined by the Free Application for Federal Student Aid.  Under this program, we gathered weekly reports from participants for academic monitoring and provided timely feedback along with dedicated tutor support.  We selected our first cohort of 17 freshmen in Fall Y+1, followed by 13 students for the entering class of Fall Y+2, 10 for Fall Y+3, and 14 for Fall Y+4, totaling 54 students.

Through Canvas~\citep{instructure2025}, we created weekly surveys for each student to submit grades, experience journals on their CS and non-CS courses, and personal challenges. Based on their reports, we administer interventions. In total, we have collected \textasciitilde 13K grade entries and \textasciitilde 7K experience journal entries in this longitudinal dataset. Although our intervention techniques somewhat reflect the viewpoint of scholarship administrators, we believe that our approach applies to the academic monitoring of student populations in CS, if not beyond.

\subsection{Research Environment Constraints}
Using cloud-based LLMs to process our dataset can breach privacy.  Although students’ and instructors’ names can be anonymized, experience journals may contain courses taught only by specific instructors during certain terms, so one can infer comments on specific instructors.  Anonymizing course names does not help, as LLMs can deduce course dependencies, which form a distinct graph, and one can identify courses based on anonymized nodes.  Journal contents about courses are also giveaways.  To prevent such leaks, we used LLMs that were hosted on local servers.  Local deployment also allows better portability and continuation of our work.

To minimize computational requirements, we limited LLMs to extracting qualitative intervention triggers or features from the journal text. We then apply lightweight techniques such as rule-based tables, decision trees, and neural networks on the extracted features to determine interventions.

\section{The CS-Guide Framework Overview}
The CS-Guide framework transforms raw student data into personalized academic interventions by combining structured grade information with unstructured journal reflections. Each week, students report their course grades and submit free-text journals describing academic and personal challenges. CS-Guide processes these inputs to identify potential issues and to recommend timely actions. The system operates through three main stages:
\begin{itemize}
    \item \textbf{Input Processing}: Students submit grades and journals weekly. The grades indicate academic performance, while journals provide narrative insight into challenges like workloads and health.
    \item \textbf{Feature Extraction}: Quantitative features are derived from structured data such as course grades and timing relative to institutional deadlines (e.g., course drops and withdrawals). Qualitative features are extracted using LLMs that identify trigger conditions (e.g., trauma, illness) from student narratives.
    \item \textbf{Intervention Prediction}: All extracted features are fed into a rule-based or machine learning system that determines which interventions are most appropriate for a student’s situation. Interventions can include reminders, academic support, mental health referrals, or contact from staff.
\end{itemize}

\begin{figure*}[ht]
  \includegraphics[width=0.75\textwidth]{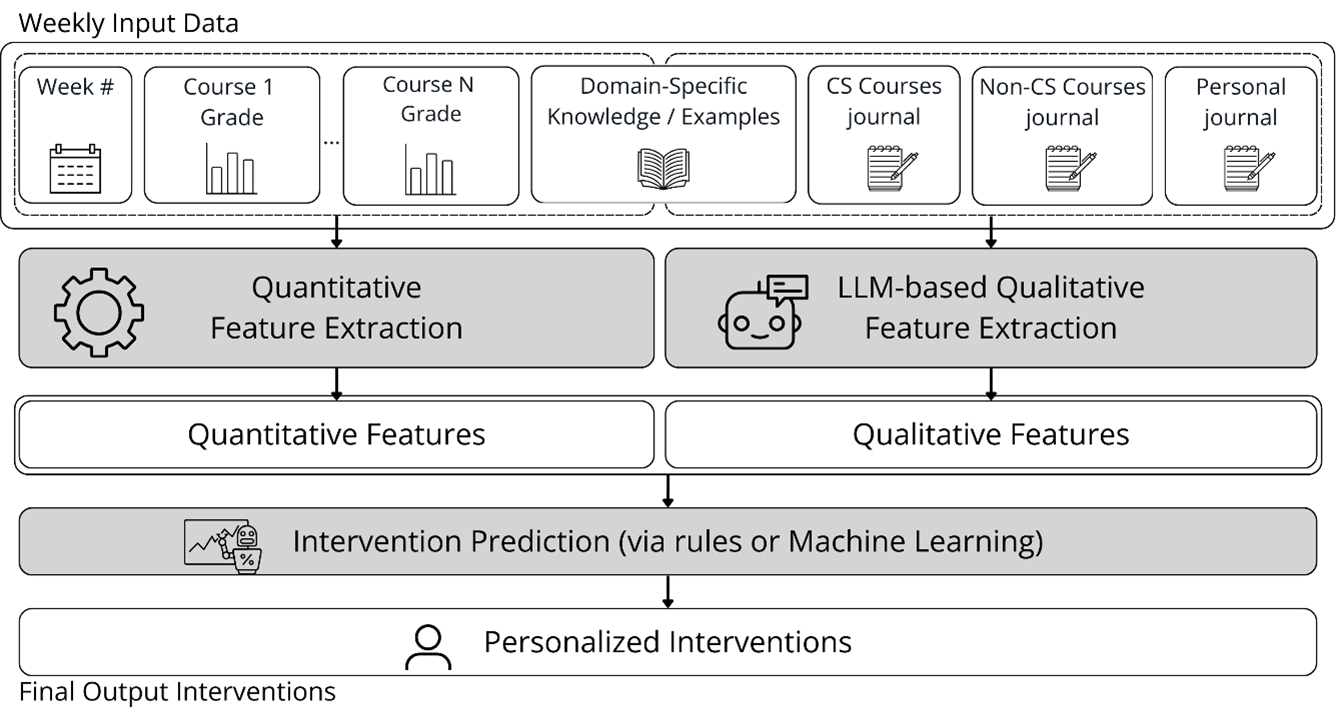}
  \caption{The CS-Guide framework. The gray boxes are our system components.}
  \Description{A flowchart illustrating our student intervention prediction framework. Raw data, including "Week Number" "Course Data" and "Student Journals," are processed through parallel "Quantitative" and "LLM Qualitative" feature extraction pipelines. Both pipelines are informed by "Domain-Specific Knowledge." The resulting quantitative and qualitative features are fed into an "Intervention Prediction" step to produce the final output: "Personalized Interventions" for each student.}
  \label{fig:architecture}
\end{figure*}

The framework is modular and designed to reflect institutional logic and constraints. Figure~\ref{fig:architecture} illustrates the architecture, where inputs (grades, journal entries, and course metadata) are transformed through these stages into actionable interventions. This layered design allows CS-Guide to scale feedback generation while remaining interpretable and extensible.

\subsection{Data Preparation and Feature Engineering}
Building CS-Guide required processing multi-year student data. To ensure the quality and utility of model inputs, we performed multiple preprocessing steps and incorporated domain knowledge into the feature extraction process.
\subsubsection{Data Alignment and Cleaning}
Due to holidays or class cancellations (e.g., extreme weather events), the report week number did not always align with the updated academic calendars and deadlines across years of the multi-year data. For consistency, report week numbers were mapped to official semester weeks, ensuring that events like midterms or drop deadlines were correctly contextualized.  

Some delimiting and non-visible characters, as well as skipped data entries, can also confuse the data alignment and processing.  We had to check empty journal columns and mismatched grades and journal entries to locate these errors.  

A non-reported grade entry may mean the course grade is not reported, not yet available, or not applicable because the course is dropped or taken with pass/fail options (e.g., seminar).  We had to use contextual information, such as grades reported from prior weeks, course numbers, and other data entries within the same week, to disambiguate them.

\subsubsection{Course Categorization and Metadata}
Since intervention strategies depend on the type of course involved, we categorized all reported courses into three categories: CS-Track Core (the CS courses required for the CS major), CS-Track STEM (the non-CS courses required for the CS major), and Other Electives. For instance, a non-required elective (Other Electives) can typically be dropped before the deadline with minimal academic or financial impact. In contrast, a required CS-Track STEM course, such as Calculus, can be taken at a community college and transferred. For CS-Track Core courses, students can be advised to retake them during the summer to allow focused effort. These distinctions, along with key academic policies like GPA thresholds and drop deadlines, enable CS-Guide to recommend interventions that are both timely and context-sensitive.
\subsection{Quantitative Features and Extraction}
Quantitative features capture observable signals of academic risk and disengagement. They fall into two categories: grade-related features (G), which reflect a student’s academic performance, and missing-report features (M), which indicate lapses in engagement with the system. These features are derived directly from structured grade reports and weekly participation patterns. For instance, a student reporting a grade below a B in CS-Track Core courses may trigger early intervention. Similarly, missing weekly reports may signal disengagement that warrants follow-ups. Table~\ref{tab:quantitative_features} summarizes the quantitative features.

\begin{table*}[ht]
\centering
\begin{tabular}{lll}
\hline
\textbf{Category} & \textbf{Feature} & \textbf{Description} \\
\hline
\multirow{7}{*}{Grade-related}
 & G1.1 & Any course grade falls below a B- in the week before the drop-deadline \\
 & G1.2 & Missing course grade in the week before the drop-deadline \\
 & G2.1 & Any course grade falls below a B- in the week before the late drop deadline \\
 & G2.2 & Missing course grade in the week before the late drop deadline \\
 & G3.1 & Any course grade below B- at the end of the semester \\
 & G3.2 & Any course grade below C- at the end of the semester \\
 & G3.3 & Any CS-Track STEM course grade below C- at the end of the semester \\
\hline
\multirow{6}{*}{Missing reports}
 & M1.1 & Missed report for 1 week \\
 & M1.2 & Missed reports 2 weeks in a row before the late drop deadline \\
 & M1.3 & Missed reports 3 weeks in a row \\
 & M1.4 & Missed 4+ consecutive weekly reports \\
 & M2.1 & Report missing during the week before the drop-deadline \\
 & M2.2 & Report missing during the week before the late drop deadline \\
\hline
\end{tabular}
\caption{Quantitative Features used in CS-Guide.}
\label{tab:quantitative_features}
\end{table*}

Quantitative feature extraction takes the structured grade reports, course categorization, and aligned week indices as inputs and computes the features directly as binary indicators. Because these inputs and outputs are well-defined (i.e., grade thresholds, number of missing reports, and the timing of missing reports), the process is straightforward and reproducible.

\subsection{Qualitative Features and Extraction}
Qualitative features capture the underlying challenges students express in their weekly journal, especially those that may not be reflected in grades. These challenges can be grouped into academic (A), health (H), personal (P), and other (O) categories. For example, a student mentioning they are behind in coursework and feel homesick would trigger academic (A2) and personal (P3) flags, respectively. CS-Guide uses an LLM to read these journals and uses curated prompt examples to detect and classify these features. These features are then encoded as binary indicators for downstream use. Table~\ref{tab:qualitative-features} lists the full set of qualitative features.  

\begin{table*}[ht]
\centering
\begin{tabular}{lll}
\hline
\textbf{Category} & \textbf{Feature} & \textbf{Description} \\
\hline
\multirow{4}{*}{Academic}
 & A1 & Needs to improve exam-taking or study skills \\
 & A2 & Falling behind or coursework too difficult \\
 & A3 & Issues with instructors \\
 & A4 & Technical issues (e.g., computer configuration problems) \\
\hline
\multirow{3}{*}{Health}
 & H1.1 & Sick for one week \\
 & H1.2 & Sick for multiple consecutive weeks \\
 & H2 & Mental health or trauma (e.g., loss of loved ones, relationship issues) \\
\hline
\multirow{6}{*}{Personal}
 & P1 & Difficulty maintaining schoolwork-life balance \\
 & P2.1 & Overcommitment (having too many hard classes or activities) \\
 & P2.2 & Job interference \\
 & P3 & Transition to college (e.g., homesickness, distractions) \\
 & P4 & Social isolation \\
 & P5 & Financial hardship \\
\hline
Other
 & O & Other uncategorized concerns \\
\hline
\end{tabular}
\caption{Qualitative features used in CS-Guide.}
\label{tab:qualitative-features}
\end{table*}

The LLM is guided with positive and negative examples in the prompt. For instance, “being away from family for the first time” is linked to (P3), while hypothetical or resolved concerns are ignored to reduce false positives.

\subsection{Intervention Triggers and Recommendations}
CS-Guide classifies interventions into four categories: (C) communication, (B) offering alternatives (Plan B), (S) providing support, and (R) referrals to other support resources. Table~\ref{tab:interventions} outlines the subtypes within each intervention category.

\begin{table*}[ht]
\centering
\begin{tabular}{lll}
\hline
\textbf{Category} & \textbf{Intervention} & \textbf{Description} \\
\hline
\multirow{4}{*}{Communication}
 & C1.1/C1.2/C1.3 & Email/text/call the student \\
 & C2 & Contact the student counselor \\
 & C3 & Contact instructors for grades \\
 & C4 & Forward student comments to the Undergraduate Curriculum Committee \\
\hline
\multirow{5}{*}{Offer alternatives}
 & B1.1 & Remind the student about drop/withdraw deadlines and minimum GPA rules \\
 & B1.2 & Suggest switching to a CS degree program with lower requirements \\
 & B2 & Recommend retroactive withdrawal \\
 & B3 & Advise taking a course elsewhere for credits that can be transferred \\
 & B4 & Recommend lowering the job workload \\
 & B5 & Suggest prioritizing schoolwork or CS-Track Core courses \\
\hline
\multirow{5}{*}{Support}
 & S1 & Offer tutor support with dedicated scholarship tutors or workshops \\
 & S2 & Study skills and exam-taking workshop \\
 & S3 & Time management workshop \\
 & S4 & College transition seminar \\
 & S5 & Social engagement event/workshop \\
\hline
\multirow{6}{*}{Referral}
 & R1 & Refer to university mentoring and tutoring or course TAs \\
 & R2 & Refer to Case Management Support \\
 & R3 & Refer to Psychological Services \\
 & R4 & Refer to the Financial Aid Office and job placement resources \\
 & R5 & Refer to the CS student counselor \\
 & R6 & Refer to the IT department \\
\hline
\end{tabular}
\caption{Intervention Categories and their Descriptions}
\label{tab:interventions}
\end{table*}

\subsection{Rule Table}
As a baseline, we handcrafted rules that mapped combinations of quantitative and qualitative triggers to specific interventions, as shown in Table~\ref{tab:rule-table}. While this framework allowed us to encode domain expertise and ensure interpretability, it quickly became unwieldy to manage as we encountered more nuanced student situations.

\begin{table*}[ht]
\centering
\begin{tabular}{lllp{7cm}}
\hline
\textbf{Feature Type} & \textbf{Category} & \textbf{Feature/Trigger} & \textbf{Interventions} \\
\hline
\multirow{10}{*}{Quantitative} 
 & \multirow{7}{*}{G (Grade related)} 
   & G1.1 & R1 \\
 &  & G1.2 & S1, interventions of G1.1 \\
 &  & G2.1 & B1.1 \\
 &  & G2.2 & C3 \\
 &  & G3.1 & B1.2, C1.1 \\
 &  & G3.2 & B2, C1.2, C1.3, C2, R5, interventions of G3.1 \\
 &  & G3.3 & B3, R5 \\
 \cline{2-4}
 & \multirow{3}{*}{M (Missing report)} 
   & M1.1 & C1.1 \\
 &  & M1.2 & C1.2, interventions of M1.1 \\
 &  & M1.3 & C1.3, C2, interventions of M1.2 \\
 &  & M1.4 & C1.1 \\
 &  & M2 & B1.1, C3, interventions of M1.3 \\
\hline
\multirow{6}{*}{Qualitative}
 & \multirow{4}{*}{A (Academic)} 
   & A1 & R1, S2 \\
 &  & A2 & R1, S1 \\
 &  & A3 & B1.1, C4 \\
 &  & A4 & R6 \\
 \cline{2-4}
 & \multirow{3}{*}{H (Health)} 
   & H1.1 & C1.1 \\
 &  & H1.2 & B1.1, R2, S1 \\
 &  & H2 & R2, R3 \\
 \cline{2-4}
 & \multirow{6}{*}{P (Personal)} 
   & P1 & R1, S3 \\
 &  & P2.1 & B1.1, B5, R1, S3 \\
 &  & P2.2 & B4, R4, intervention for P2.1 \\
 &  & P3 & R1, S4 \\
 &  & P4 & S5 \\
 &  & P5 & R2, R4 \\
 \cline{2-4}
 & O (Other) & O & R5 \\
\hline
\end{tabular}
\caption{Summary of intervention rules for each trigger.}
\label{tab:rule-table}
\end{table*}

Many triggers overlapped in both logic and timing. For instance, missing two weekly reports (M1.2) is a superset of missing one (M1.1), meaning interventions can be cumulative: M1.2 adds texting (C1.2) to the emailing (C1.1) of M1.1. After four missed reports (M1.4), we de-escalate to weekly emails and stop further outreach, especially after late drop deadlines. Two features with different interventions can be triggered at the same time.  For instance, when the fourth missing report (M1.4) occurs the week before the drop deadline (M2), we carry out interventions for both, unless the classes are canceled.  

Triggers and interventions can be based on the recent past.  For instance, if the student misses 4+ reports (M1.4), we stop escalating the contact efforts and return to weekly emails (C1.1). The escalation also stops after the late drop deadline.  As another example, a student only needs to attend a specific workshop once per semester (S2-S4).

Certain combinations can also generate conflicting recommendations. A student requesting help with exam-taking skills (A1) but with solid grades (not G1.1 and not G1.2) may benefit from a study-skill workshop (S1). However, referring them to tutors (R1) could be less effective. We also cannot rely solely on grades to be posted in a timely manner. In another case, a student struggling with overcommitment (P2.1) and social isolation (P4) may lack the time or willingness to attend social support events (S5). Health and academic issues can also combine in problematic ways; for instance, a student who is ill (H1.*) and academically behind (A2) may be too burdened to follow through on offered help (S1, R1).

Furthermore, the impact of certain events varies over time. Mid-semester course drops generally do not affect academic standing (G2.*), whereas poor final grades (G3.*) may trigger longer-term interventions. 

As the number of corner cases increased, we realized that our intervention logic had become a high-dimensional Venn diagram of trigger-intervention interactions. Simple modifications to a rule could inadvertently trigger cascading effects on unrelated interventions. To address this complexity and improve scalability, we introduced machine learning models to automate intervention prediction.

\subsection{Machine Learning Techniques}
We experimented with decision tree-based methods, such as categorization and regression trees (CART)~\citep{loh2011classification}, random forests~\citep{breiman2001random}, and XGBoost~\citep{chen2016xgboost}. These models treat each intervention as a separate prediction task: for every intervention type, a binary classifier is trained using all extracted quantitative and qualitative features. This allows the model to learn associations between features and interventions from data rather than relying solely on hand-engineered rules.

In addition to tree-based models, we implemented a multilayer perceptron (MLP) neural network~\citep{murtagh1991multilayer} that jointly predicts all applicable interventions. The MLP takes the full set of features as input and outputs a multi-label classification over all interventions. Its capacity to model nonlinear feature interactions makes it well-suited for capturing complex trigger combinations and subtle contextual nuances.

Together, these machine learning techniques supplement our rule-based system, offering a data-driven pathway to generalize from historical student cases and potentially improve the accuracy and personalization of interventions over time.

\subsection{Implementation Details}
CS-Guide is designed as a lightweight, local system that integrates data processing, language modeling, and machine learning pipelines. Our prototype was implemented in Python 3.1, using the openpyxl (3.1.5) and pandas (2.2.3) libraries for spreadsheet parsing and data manipulation. For natural language understanding, we deployed Llama 3.1:8B~\citep{grattafiori2024llama} locally using Ollama 0.4.8 on an NVIDIA RTXTM A6000 GPU. This allowed us to maintain data privacy and control over model behavior.

\begin{figure*}[ht]
  \centering
  \includegraphics[width=0.7\linewidth]{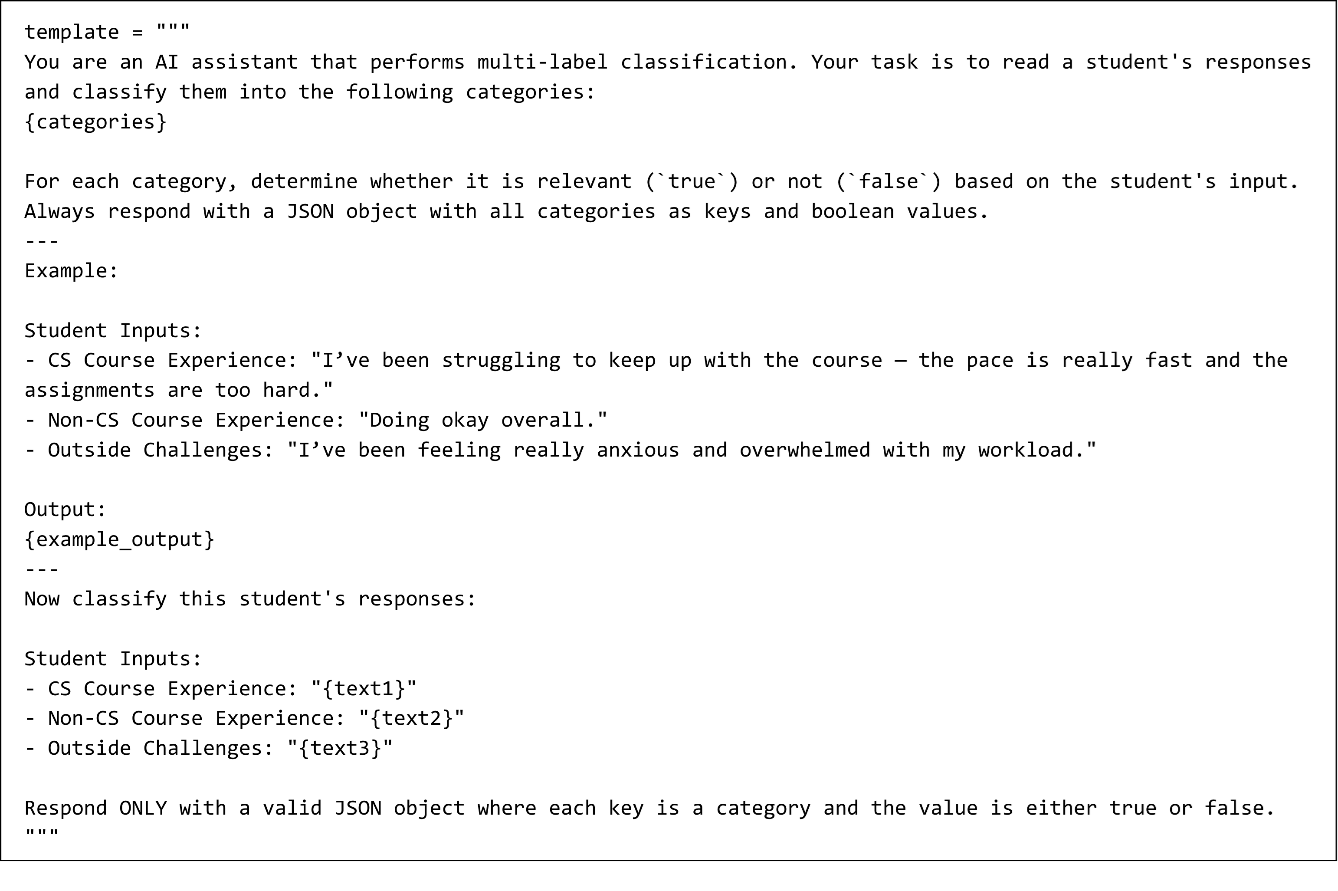}
  \caption{Simplified prompt template to extract qualitative features from student experience journals.}
  \Description{[TEXT] Simplified Llama prompt example to extract qualitative features from student experience journals.}
  \label{fig:prompt}
\end{figure*}

The pipeline begins with preprocessing, which aligns weekly calendar dates and cleans up raw student data. Quantitative features such as grades and report compliance are then heuristically extracted. To extract qualitative features from students' weekly journal reflections, we use a structured prompt with example-based few-shot learning (Figure~\ref{fig:prompt}) and set the model temperature to 0.1 to minimize variability and promote reproducibility. The LLM returns structured JSON feature annotations for downstream use.

For intervention prediction, we implemented several methods: a rule-based baseline, three variants of decision trees (CARTs, random forests, and XGBoost), and an MLP neural network. The random forest model used a maximum depth of 15 and a minimum rule frequency of 0.1. The CART models were configured with \texttt{max\_depth} = 15, \texttt{min\_samples\_split} = 15, \texttt{min\_samples\_leaf} = 5, and \texttt{class\_weight} = "balanced", with additional constraints of \texttt{max\_rules\_per\_target} = 10 and \texttt{min\_samples\_for\_rule} = 10. The MLP used a 256--128--64 architecture with dropout (0.3), L2 regularization (0.001), batch normalization, and was trained for 100 epochs with a learning rate of 0.001; all other parameters were set to defaults. These models ingest weekly context (week number), quantitative features, and LLM-derived qualitative features to predict applicable interventions. Postprocessing routines verify results against ground truth, generate statistical summaries, and track LLM reasoning outputs to support debugging and explainability.

\section{Evaluation}
\textbf{Datasets for Training and Testing}: From Fall Y+1 to Spring Y+4, we collected \textasciitilde 20K data entries (\textasciitilde 13K grade-related and \textasciitilde 7K experience journal entries), yielding \textasciitilde 170K intermediary data points consisting of contextual features and output interventions.  However, due to the need for extensive manual inspection of 100+ raw data entries and 900+ intermediary and output items per student per semester, we processed only \textasciitilde 3,400 data entries and \textasciitilde 28,000 intermediary and output entries from Fall Y+1 and Spring Y+2 for training (\textasciitilde 17\% of the full dataset).

For testing, we randomly selected three first-year students from each of the year Y+2, Y+3, and Y+4 cohorts, using \textasciitilde 1,600 data entries from Fall and Spring Y+2 to Fall and Spring Y+4. The selected training and testing data entries were manually labeled and used as the ground truth for evaluating predictions.

\textbf{Metrics}: We evaluate the performance of CS-Guide using four key metrics. Accuracy reflects the overall correctness of predictions, calculated as the ratio of true positives and true negatives to total predictions. We evaluated the model's predictions against the manually labeled ground truth data. Precision measures the proportion of correct positive predictions, while Recall focuses on how well the model identifies actual positives. The F1 Score combines precision and recall, offering a combined metric. These metrics provide a broader view of model performance.
\begin{description}
  \item[Accuracy:] $(TP + TN) / (TP + TN + FP + FN)$
  \item[Precision:] $TP / (TP + FP)$
  \item[Recall:] $TP / (TP + FN)$
  \item[F1 score:] $2 \cdot (\text{Precision} \cdot \text{Recall}) / (\text{Precision} + \text{Recall})$
\end{description}
Here, $TP$, $TN$, $FP$, and $FN$ denote true positives, true negatives, false positives, and false negatives, respectively.

\subsection{Early Experiences with LLMs}
Due to time constraints, we evaluated the system using only Llama 3.1, although we acknowledge that newer, more advanced models may offer similar or even better performance in analyzing student journals. This may be a direction for future assessment. To date, our early experiments highlight the challenges of natural language processing in this analysis.
\begin{itemize}
  \item \emph{Lack of freshmen context}: The LLM lacked context about students being college freshmen and was unable to identify common adjustment issues (e.g., homesickness, oversleeping) as college transition issues (P3).
  \item \emph{Context misinterpretation}: Terms like “too much work” were often interpreted as job-related (P2.2) instead of schoolwork. “Doing projects” was occasionally misread as extracurricular involvement.
  \item \emph{Overdiagnosis}: Common phrases like “feeling anxious” during exam week were sometimes flagged as mental illness (H2), triggering referrals (R2, R3).
  \item \emph{Scheduling confusion}: Future course planning issues were sometimes interpreted as work-life imbalance (P1).
  \item \emph{Literalism}: Plain reporting of school life (e.g., “having lots of schoolwork”) was mistaken for problems, leading to potentially unneeded interventions.  Having “free time” early in a semester can be flagged as not maintaining work-life balance (P1) and trigger referrals and support (R1, S3).
  \item \emph{Self-intervention}: The LLM would still trigger interventions even when a student reported having taken or was planning to take the same interventions, such as meeting with a counselor or dropping a class.
  \item \emph{Hypotheticals as problems}: Pessimistic comments about future courses were interpreted as current issues.
  \item \emph{Typo sensitivity}: Minor errors like miswritten course codes were treated as “other issues” (O).
  \item \emph{Odd classifications}: For example, “can’t sleep due to a death in the family” was classified as work-life imbalance (P1) rather than trauma-related (H2).
  \item \emph{Hallucinations}: In rare cases, the LLM interpreted succinct journal entries such as “no challenges” as work-life balance (P1), college transition (P3), or other issues (O). 
\end{itemize}
To improve performance, we refined our prompts with clarifying examples for college transitions and counselor-related issues and incorporated additional rules to consider grades and illness when recommending interventions. We recognize that many challenges are inherent to NLP, beyond our scope. 

After refining the rules and prompts from the first pass, our second pass led to improved performance, with other issues:
\begin{itemize}
    \item \emph{Boolean triggers} may be problematic.  For example, a vague \lq lack of motivation' might signal mental health concerns (H2), flagged by the LLM, but it is not enough to recommend psychological services (R3) without stronger evidence.
    \item \emph{Course prefixes not enough}: Although course prefixes were included to assist classification, the LLM occasionally missed domain-specific signals (e.g., chemistry course issues classified as “other”). 
\end{itemize}
Further LLM refinement remains as future work.

\subsection{CS-Guide Performance}
Table~\ref{tab:performance-metrics} presents the system’s overall performance.  Quantitative feature extraction was fast (\textasciitilde 7 seconds) and accurate since this step involves simple accounting of grades and missed reports. Qualitative feature extraction required more time (\textasciitilde 40 minutes of LLM processing) and achieved \textasciitilde 93-95\% across all metrics. Intervention prediction using the LLM combined with rules or machine learning reached 95–97\% accuracy, with training completed in under one minute and inference in \textasciitilde 3 seconds.

\begin{table*}[ht]
\centering
\begin{tabular}{llcccc}
\hline
\textbf{Category} & \textbf{Method} & \textbf{Accuracy} & \textbf{Precision} & \textbf{Recall} & \textbf{F1} \\
\hline
\multirow{1}{*}{Quantitative Feature Extraction} 
 & Python & 100\% & 100\% & 100\% & 100\% \\
\hline
\multirow{1}{*}{Qualitative Feature Extraction} 
 & LLM & 93\% & 95\% & 93\% & 94\% \\
 \hline
\multirow{5}{*}{Intervention Predictions} 
 & LLM + Rule-Based & 96\% & 96\% & 96\% & 96\% \\
 & LLM + CARTs & 95\% & 96\% & 95\% & 95\% \\
 & LLM + XGBoost & 97\% & 97\% & 97\% & 97\% \\
 & LLM + Random Forests & 96\% & 96\% & 96\% & 96\% \\
 & LLM + MLP & 96\% & 96\% & 96\% & 96\% \\
\hline
\end{tabular}
\caption{CS-Guide performance metrics.}
\label{tab:performance-metrics}
\end{table*}

With a small training set and modest hardware requirements, the system can produce consistent and accurate intervention predictions using both rules and learned models. The confidence intervals at the 90\% level were within 1\% for all metrics, indicating that performance estimates are statistically stable despite the relatively small dataset. Although there are alternative optimizations and system designs, these numbers suggest that the feasibility of CS-Guide is promising.

\section{Lessons Learned}
In developing CS-Guide, we encountered a few challenges that shaped both the design of the system and our understanding of its use in practice.

\textbf{Verification highlights the fragility of rule-based systems}. Debugging rule-based interventions required time-intensive verification, as each rule update risked breaking rule dependencies elsewhere. This highlighted the fragility of handcrafted logic and the value of machine learning for scalability and dynamic adaptation without exhaustive revalidation.

\textbf{Timely data availability and reliability shape intervention quality}. Delayed grade posting and system-projected grades complicate interventions. A few large introductory courses did not post grades before the drop deadline, due to assignment delays caused by SSH configuration issues. In response, we worked with the systems group to deploy a browser-based SSH service in the subsequent semesters. In other cases, instructors revealed that Canvas posts the projected course grade with unsubmitted assignments (even past the submission deadlines) set to the average scores of submitted ones.  Thus, missed assignments do not get zeroes.  This practice left students with inflated grades and delayed interventions. We reported this issue to the university IT department.

\textbf{Lag in grade reporting}.  Reported grades reflected only the last posted values, meaning that they may not reflect a student’s current performance.  Students sometimes reported doing well on exams before official grades were entered, leading to mismatches between the system’s triggers and the actual situation. This suggests the need for future “qualitative override” triggers that incorporate journal entries more dynamically.

\textbf{Missing reports may signal deeper issues}. We found that missing reports, especially during the COVID-19 pandemic, were strong signals of student distress. Proactive outreach in such cases led to timely interventions, including medical withdrawals and changing to another computer science degree with fewer requirements. More broadly, we discovered that absences and non-responses should not be treated as noise, but as informative features.

\textbf{Ethical considerations must guide interventions}. Recommendations to drop a course may have significant degree-progress and financial consequences, potentially discouraging students from continuing in the major. Advising systems must frame interventions as options with clarified trade-offs. 

\textbf{Corner cases are inevitable}. Despite extensive rules, new exceptions emerged on the academic and employment sides. For example, some scholarships (including substantial state-supported scholarships common among our sample) prohibit dropping below a full course load. Meanwhile, certain employment contracts impose severe penalties for reducing hours. In these cases, interventions that might normally be appropriate became harmful. This experience underscored the importance of designing CS-Guide to be flexible and amendable to new rules as unexpected scenarios arise.

\section{Limitations and Future Work}
Our study reflects the perspective of CS scholarship administrators and faculty rather than professional academic advising or personal counseling staff. While this lens shapes the current design of CS-Guide, we believe the system can be tailored for advising contexts more broadly and adapted to different student populations beyond computer science.

At present, our analysis is limited to data from the first academic year (Fall Y+1 and Spring Y+2). Some of the triggers and interventions we developed may not generalize to later stages of students’ academic journeys. As we expand our training dataset across multiple academic years and cohorts, we anticipate refining existing rules and adding new triggers and interventions that capture the evolving challenges students face over time.

Another limitation is the reliance on pretrained models not tailored for this qualitative feature extraction. While hallucinations rarely arise in our initial experiments, their occurrences may increase as we scale to more data and additional models. Exploring a broader set of LLMs, machine learning techniques, and data-processing pipelines will allow us to benchmark performance and assess robustness. However, our experience also highlighted that verification would remain a significant bottleneck, requiring careful design of human-in-the-loop processes.

Future work will focus on extending our feature design. For example, Boolean qualitative triggers such as “mental health issues (H2)” could be subdivided into finer categories to distinguish between situations that can be resolved via peer support (e.g., a stress management workshop) versus those that require clinical referrals. Similarly, for lags in grade postings, we plan to implement qualitative override triggers informed by students’ journals to adjust interventions to be more responsive.

We also plan to enhance contextual richness. Extracting per-course journals could enable more targeted interventions, while incorporating course metadata (e.g., titles, credit hours, grading options) could help students better prioritize their efforts and provide LLMs with richer context.

Finally, we plan to investigate strategies for privacy-preserving data sharing. By redacting sensitive information, we aim to make our dataset accessible to the broader CS education research community, enabling replication, cross-institutional comparison, and further refinement of advising-support systems.

\section{Related Work}
Since LLMs have been applied to many areas of education, we focus here on systems related to academic counseling and student support. A growing number of platforms integrate chatbots with institutional knowledge bases to guide course selection and academic planning~\citep{abdelhamid2025advisely, newton2025AIDrivenAcademic}. Recent extensions include AI-driven advising and LLM-powered chatbots designed for personalized student guidance~\citep{aguila2024large, ismail2025fine, khader2025smart, lekan2023ai}. These systems demonstrate the potential of conversational agents to deliver accessible advising at scale~\citep{igualde2024guidelines, radhakrishnan2023use, tamascelli2025academic}. However, most remain agnostic to majors, focusing on navigation of university requirements at the time scale of academic terms. CS-Guide complements this line of work by emphasizing fine-grained, discipline-specific monitoring and proactive outreach.

Academic monitoring systems more broadly have become common in postsecondary institutions as part of student success and retention initiatives~\citep{parker2025large, radford2018improving, wise2021subversive, zhang2024dr}. These tools typically identify students at risk based on course performance signals or survey data and are not discipline-specific. More recent work extends this approach by developing AI-driven and LLM-powered advising systems that adapt interventions to individual students and contexts~\citep{newton2025AIDrivenAcademic,tamascelli2025academic}. CS-Guide contributes to this trajectory by addressing the complexities of overlapping academic, personal, and health-related signals and by coordinating interventions across multiple stakeholders (students, advisors, and instructors).

Other related domains include tools that provide targeted feedback to students. These tools span work on reducing procrastination and supporting self-regulation~\citep{bhattacharjee2024understanding}, knowledge tracing to predict learner outcomes~\citep{newton2025AIDrivenAcademic}, and automated feedback or grading systems~\citep{jia2024llm, letteri2024exploring, meyer2024using}. In parallel, LLMs have been explored in mental health support for higher education, pointing to the potential of natural language models in addressing sensitive well-being issues.

Finally, there is a substantial body of research on predicting student performance using machine learning approaches~\citep{ahmed2024student, chen2025machine, wang2025machine, wayesa4656835analysis}. These works primarily focus on outcome prediction from structured data such as grades and activity logs, whereas CS-Guide integrates both quantitative (e.g., grades, missed reports) and qualitative (e.g., journals) signals to not only predict risks but also recommend interventions.

\section{Summary and Contributions}
We introduced CS-Guide, a framework that leverages LLMs to provide timely, scalable, and discipline-specific academic monitoring for computer science students.  This work has the following contributions:
\begin{itemize}
    \item \textbf{A longitudinal dataset}: We collected a four-year academic monitoring dataset consisting of \textasciitilde 20,000 weekly student grades and experience journals.
    \item \textbf{Feature and intervention identifications}: We identified 13 quantitative features and 14 qualitative features in 5 categories, as well as 24 interventions in 4 categories in the domain of CS academic monitoring.
    \item \textbf{The CS-Guide framework}: We designed and implemented CS-Guide, which integrates quantitative features (e.g., grades, attendance) and qualitative features extracted from student journals using LLMs, to generate academic interventions.
    \item \textbf{Evaluation across cohorts}: We evaluated CS-Guide on the entering academic year of the first cohort, and tested its generalizability using three additional freshman cohorts, demonstrating the feasibility and scalability of the approach.
    \item \textbf{Insights into LLM behavior}: We analyzed common misinterpretations of student journals by Llama, documenting both its limitations and the lessons we learned from adapting LLMs for academic monitoring.
\end{itemize}
	
Overall, our findings demonstrate the feasibility of combining structured academic records with LLM-based journal analysis to support proactive, data-driven student advising. With continued data collection and refinement, CS-Guide has the potential to evolve into a robust, generalizable framework for academic monitoring in computer science and beyond.

\bibliographystyle{ACM-Reference-Format}
\bibliography{custom}










\end{document}